%% file: IV2010.tex
\documentclass[10pt, conference]{IEEEtran}

\usepackage{times}
\usepackage{graphicx}
\usepackage{url}
\usepackage{amssymb}
\usepackage{cite}

\usepackage{fancyhdr}
\begin{document} 

\sloppy

\title{\Large\bf Combining Visual Analytics and Content Based Data Retrieval Technology for Efficient Data Analysis\\
{\ }\\
\footnotesize{IEEE Copyright - http://ieeexplore.ieee.org/xpls/abs\_all.jsp?arnumber=5571357}}

\author{Jose Rodrigues, Luciana Romani, Agma Traina, Caetano Traina\\
       Instituto de Ciencias Matematicas e de Computacao - Universidade de Sao Paulo \\
       Universidade Federal de Sao Carlos - Campus Sorocaba\\
       Embrapa Agriculture Informatics - Campinas\\
       $[$junio,agma,caetano$]$@icmc.usp.br, luciana@cnptia.embrapa.br}

\maketitle
\thispagestyle{fancy}

\begin{abstract}
One of the most useful techniques to help visual data analysis systems is interactive filtering (brushing). However, visualization techniques often suffer from overlap of graphical items and multiple attributes complexity, making visual selection inefficient. In these situations, the benefits of data visualization are not fully observable because the graphical items do not pop up as comprehensive patterns. In this work we propose the use of content-based data retrieval technology combined with visual analytics. The idea is to use the similarity query functionalities provided by metric space systems in order to select regions of the data domain according to user-guidance and interests. After that, the data found in such regions feed multiple visualization workspaces so that the user can inspect the correspondent datasets. Our experiments showed that the methodology can break the visual analysis process into smaller problems (views) and that the views hold the expectations of the analyst according to his/her similarity query selection, improving data perception and analytical possibilities. Our contribution introduces a principle that can be used in all sorts of visualization techniques and systems, this principle can be extended with different kinds of integration visualization-metric-space, and with different metrics, expanding the possibilities of visual data analysis in aspects such as semantics and scalability.
\end{abstract}


\section{Introduction}
\label{sec:introduction}

\noindent{The concept of metric space is the main solution for the problematic of managing data similarity. According to this methodology, a multivariate data instance (a record) refers to the descriptive characteristics of a single entity in a specific domain. Having a set with multiple data items (a dataset), and with the aid of a distance function (or similarity measure), it becomes possible to determine an {\it order relation} within the dataset. Once that relation is set, one can perform similarity queries over the data and have results that reflect certain properties of the entities; the reflected properties will be defined according to the very data and according to the distance function being used. This kind of functionality is also known as {\it content-based data retrieval} and has a large scope of application, from bioinformatics to complex-data storage systems. In this work, benefiting from this methodology, we combine metric space techniques and visualization techniques in order to discover data patterns in a multiple views environment.}

Similarity queries, the goal of metric space systems, work by searching a dataset and tracking for the elements that, according to a distance function, are similar to a given data element, named {\it query center}, within a given scope. The result of a similarity query is a subset of elements, as much as possible, similar to the query center. In an analogical view, the query center and its scope of data retrieval resembles the idea of a region in space; in such region the center is the query center surrounded by neighboring data elements. This analogy is the basic idea of our methodology:

\begin{enumerate}
	\item{first, the user must refine the dataset with the aid of similarity queries;}
	\item{the result sets produced by these queries are the focus of visualization workspaces enriched with multiple visualization techniques.}
\end{enumerate}
 
We propose an environment that allows the user to browse a dataset in its tabular format, visualize such data, select query centers, perform similarity queries and have the results of the queries drawn into visualization workspaces that co-exist in the system. Our contributions aim at inserting intelligence into the problematic of precisely and semantically filtering a dataset into complementary visualizations. Our future goals are to foster works on scalability and semantical visual data analysis.

Visualization techniques in general are designed to work with the entire dataset, providing an initial overview of the data, followed by interactive brushing (selection). However, what if the dataset is too complex (with too many attributes), voluminous and/or heterogeneous, making visualizations inefficient due to overlapping of graphical items and due to cluttering of visual information across different visualization techniques? In the context of these issues we aim at answering the following questions:

\begin{itemize}
	\item{How to filter out data in situations where interactive brushing is not efficient due to overlapping of graphical elements or due to a high number of data attributes?}
	\item{How to add semantics in the visual data analysis process, providing better visual feedback in the context of a specific application domain?}
	\item{How to quickly make sense of metric spaces in a visual manner?}
\end{itemize}

Our results show that the use of a metric space can significantly improve the quality of visualizations, enabling us to answer those questions. In our work, we proceed by precisely configuring what data items are being displayed and how these items are arranged in multiple workspaces.

In order to test our hypotheses, we have built a system named Metric Space Platform (MetricSPlat) (\url{http://gbdi.icmc.usp.br/~junio/MetricSPlat/index.htm}). MetricSPlat is a system that integrates the components that define a metric space so that content-based data retrieval can be accomplished in a visual interactive environment. It uses plug-in features and coding frameworks to permit the fast prototyping and testing of features extraction techniques, distance functions and metric structures. The ensemble of the system joins research concepts from database and Visual Analytics in a novel manner, bringing an original contribution to both fields. The paper has further five sections: related works, concepts, proposed methodology, experiments and conclusions.

\section{Related works}
\label{sec:related_works}

\noindent{According to Rundensteiner {\it et. al.} \cite{RundensteinerWYD02}, conventional multivariate visualization techniques present limitations that result in a display with an unacceptable level of clutter. In fact, most visualization techniques suffer with data overlapping and with datasets whose elements tend to spread over the data domain. Therefore, such visualizations provide views with a reduced number of noticeable differences.}



The most used alternative to troubleshoot the anomalies of data visualization is the interactive filtering principle that, according to Keim \cite{14}, claims that ``in exploring large datasets, it is important to interactively partition the dataset into segments and focus on interesting subsets''. Following that principle, many other authors developed tools aiming at the interactive filtering goal, as the Polaris system \cite{Chris2008} and dynamic querying tools \cite{Hochheiser2004}. Selective visualization is fundamental since it enriches the user participation during the visualization process, allowing users to focus on partitions of more interest by constantly redefining the view in order to characterize the data under analysis.


Gerken {\it et. al.} \cite{Reiterer2009} make an extensive review of visual information seeking systems. They state four conditions for the design of efficient such systems: (1) to support various ways of formulating an information need, (2) to integrate analytical and browsing-oriented ways of exploration, (3) to provide views on different dimensions of the information space, and (4) to make search a pleasurable experience. In this work, we consider these recommendations by proposing a methodology that puts together the analytical power of metric spaces and that is flexible in a multiple-ways query interface.

Following the idea of metric spaces and visual data analysis, Hiroike {\it et. al.} \cite{HiroikeMSM99} developed a system that presents the results of a content-based image retrieval system in a scatter diagram. In the diagram, image thumbnails have their size determined by their similarity to an image used as the query center. The arrangement of the images is determined by the values found in their respective features vectors. The system is introduced as an interface for content-based image retrieval.

\section{Basic concepts}
\label{sec:basics}

\noindent{This section reviews concepts of metric space and content-based data retrieval, part of the techniques used in our work.}

\subsection{Multivariate data}

\noindent{The first thing in order to index a set of complex data is to have it in an appropriate multivariate format, popularly called tabular form. In many domains, such as image and video, this step demands a process of {\it features extraction}. That is, one must translate the data into a numerical representation that corresponds to a vector $x=\{x_0,x_1,...,x_{n-1}\}$ of $n$ representative numbers intrinsic to the original data. Classic examples, in the case of images, are the color histogram \cite{Felipe2005a} and the coefficients achieved with the Fourier transform \cite{Zhang2001b}. In other domains, data must simply be collected in order to represent the instances of observed events. That is, the features extraction occurs according to a previously established model, as for example, in commercial transactions, census data and remote sensing data.
Along this text, we refer to features extraction in general as a function $f:D \rightarrow \mathfrak{D}$, where $D$ is a domain of specific data, e.g. images, and $\mathfrak{D} \subset R^n$ is an $n$-dimensional features space.}\\

\subsection{Distance function}

\noindent{The second step in order to define a metric space is to establish a similarity measure, or distance function, among the vectors of numbers extracted from the data objects. A trivial way to do this is to consider each numerical feature as an $n$-dimensional coordinate and calculate the Euclidian distance among the vectors. Other examples of distance functions are the City Block and the Minkowisk distances \cite{Aggarwal2001e}. A distance of particular interest in this work is the weighted Minkowisk distance \cite{Randall97}, defined as follows:}

\begin{equation}
	\delta_{Minkowski}(o_i,o_j) = \sqrt[p]{ \sum_{i=1}^n w_i\left(x_i - y_i\right)^p }
	\label{eq:MinkowskiPond}
\end{equation}

Where $o_i=\{x_0,x_1,...,x_{n-1}\}$ and $o_j=\{y_0,y_1,...,y_{n-1}\}$ are vectors with $n$ representative numerical features, $0 \leq i < n $, $o_i$ and $o_j \in \mathfrak{D}$, the domain of elements to be indexed. The vector $w=\{w_0,w_1,...,w_{n-1}\}$ is a vector of $n$ weights.

The use of different metric distances allows not only having a diversity of scopes around the query center in similarity queries, it also permits to weight specific dimensions, adding semantical interest into the query execution. These two parametrization possibilities correspond to a guided distortion of the space according to the needs of the analyst, who may decide about which dimensions are more relevant in a specific domain.\\

\subsection{Metric space}

\noindent{Once feature vectors and a distance function are specified, a metric space is established. A metric space refers to a set in which the notion of distance among its elements is well-defined. Formally, a metric space is a pair $M = <\mathfrak{D},\delta()>$, where $\mathfrak{D}$ is the domain of the elements to be indexed and $\delta:\mathfrak{D} \times \mathfrak{D} \rightarrow \Re^+$ is a function that associates a distance to any pair $o_i, o_j \in \mathfrak{D}$. Moreover, given three elements $o_i, o_j$ and $o_k \in \mathfrak{D}$, the pair $M = <\mathfrak{D}, \delta()>$ is named metric space whenever the function $\delta()$ satisfies the following axioms:}
	
\begin{enumerate}
\item {\bf Symmetry}: $\delta(o_i, o_j) = \delta(o_j, o_i)$
\item {\bf Non negativity}: $0 < \delta(o_i, o_j) < \infty $ if $o_i \neq o_j$ and $\delta(o_i, o_i) = 0$
\item {\bf Triangular inequality}: $ \delta(o_i, o_j) \leq \delta(o_i, o_k) + \delta(o_k, o_j)$ \label{item:desTriang}
\end{enumerate}

A metric space, as stated by its name, embeds numerical vectors inside a space in which the distances among objects are kept coherent following the properties of the metric axioms. The spatial analogy is what makes the concept so useful for humans, who can intuitively operate in such a space through the notion of similarity.\\

\subsection{Similarity queries}

\noindent{Over a metric space, it becomes possible to perform similarity queries. That is, given an element of interest -- the center of the query -- what are the elements of the dataset with smaller distances (higher similarities) to this element. The most well-know similarity query is the Nearest-Neighbor:}\\

\noindent{{\bf Definition 1(Nearest-Neighbor query)}: Given a query object $o_q$ represented by its features vector $f(o_q)$, and the set of data elements $D$, the nearest neighbor is the element of $D$ such that $NNQuery(o_q)=\{o_n \in D | \forall \ o_i \in D, \delta(f(o_q), f(o_n)) \leq \delta(f(o_q), f(o_i))\}$. An example of a nearest neighbor query is: ``find the enterprise record in $D$, which is the most similar to enterprise record $o_q$''.}

The extrapolation of definition 1 for $k$ nearest neighbors, $k \geq 1$, is straight and determines what is known as {\it K Nearest Neighbors} query (KNNQuery). Formally, a $KNNQuery(o_q,k)$ generates an ordered list in which the $(n-1)$-th-element is closer to $o_q$, or at the same distance, than the $n$-th-element, $2\leq n \leq k$.

\section{Combining metric spaces and visualization techniques}
\label{sec:system}

\noindent{In this section we describe the MetricSPlat system and how it defines a framework that supports content-based data retrieval combined with visual analytics.}

\subsection{Overview of the Metric Space Platform (MetricSPlat)}

\noindent{In order to versatilely use features extraction, MetricSPlat has an embedded dataset facility. This way, the system can perform similarity queries over any given set of extracted features provided in a simple tabular format. The format requires a table in which the {\it n} first fields correspond to the numerical features. Besides the features, it is necessary to have an extra field named {\it COD}, an integer indexing counter, as the last field (left to right order) of the dataset. No database installation is necessary.}\\ 

\noindent{{\bf Distance functions and metric structures}}\\
\noindent{In MetricSPlat, we benefit from an Application Programming Interface (API) named {\it Arboretum} -- \url{http://www.gbdi.icmc.usp.br}. This API defines a set of software interfaces to ease the development of distance functions and metric structures supporting the execution of similarity queries. In order to make MetricSPlat easily extensible to new functionalities, we have developed it around a Dynamic Linking Library (DLL) framework in compliance to the Arboretum API. Like so, MetricSPlat automatically recognizes new dll files making them ready for use with no recompilation.}\\

\noindent{{\bf Data visualization}}\\
\noindent{MetricSPlat also offers a set of visualization techniques to permit the visual inspection of the dataset under analysis and of the result set produced by similarity queries. When the results of similarity queries are under investigation, the system builds dedicated workspaces that support multi-modal visual analysis. After the execution of a nearest-neighbor query, a dedicated workspace is created loaded with the correspondent result set. Each workspace is initially presented using the FastMap spatial projection technique.}

This technique is based on the dimensionality reduction algorithm FastMap \cite{Faloutsos1995}, which is used to reduce the dimensionality of the features dataset. FastMap is an instance of multidimensional projection, that is, a function $m: \mathfrak{D} \rightarrow \Re^3$ whose goal is to minimize the sum of the distances differences between functions $\delta$ (original space) and the Euclidean distance $d$ (projection space). That is:

\begin{equation} Argmin_{m}({\sum_{i,j}|\delta\{(o_i,o_j)\}-d\{(m(o_i),m(o_j)\}|)}
\end{equation}

Where $o_i$ and $o_j \in \mathfrak{D}$.

Once the dataset has its dimensionality reduced to $3$ dimensions, it becomes possible to have it in a $3$-dimensional plot space. In such representation, one can observe clusters, exceptions and, more important, the spatial distance-based placement of the data. This possibility corresponds to the visual observation of a metric space allowing to see how a similarity query works and what are its results, as it can be seen in figure \ref{fig:visualization}.

\begin{center}
\includegraphics[width=\columnwidth]{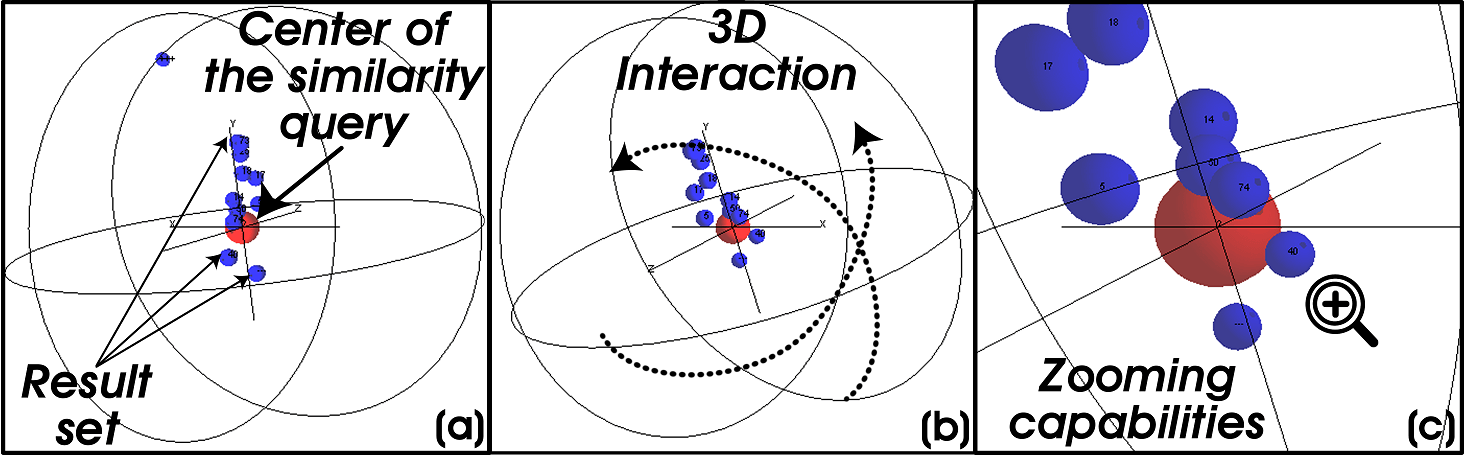}
\makeatletter\def\@captype{figure}\makeatother
\caption{Visualization of a result set in a $3$-dimensional environment. (a) The query center as the origin of the projection. (b) 3D interaction. (c) Zooming capabilities.}
\label{fig:visualization}
\end{center}

Besides FastMap for spatial data projection, MetricSPlat provides well-known techniques for visualizing the extracted features that define the search space: Parallel Coordinates \cite{Inselberg90}, Scatter Plots, Table Lens \cite{Rao94}, and Star coordinates \cite{Kandogan2000}.


\subsection{MetricSPlat systematization}
\label{subsec:systematization}
\begin{figure*}
 \begin{center}
  \scalebox{0.4}{
    \includegraphics*{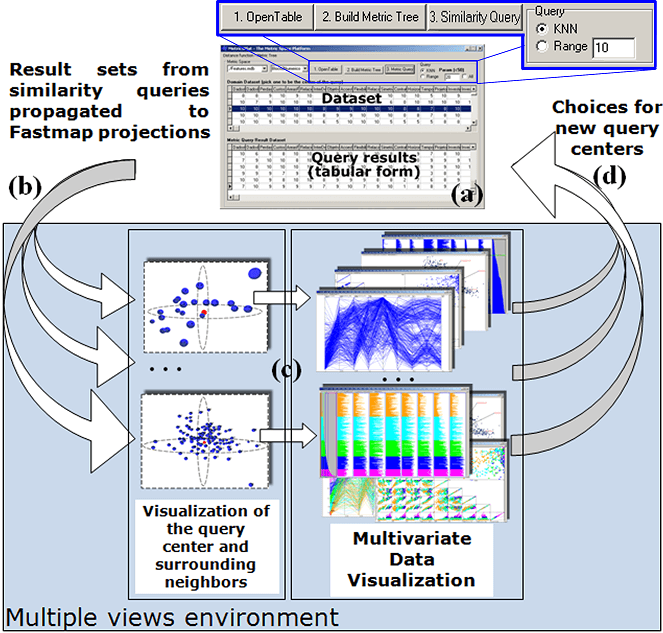}
  }
 \end{center}
 \caption{
The MetricSPlat multiple views environment. (a) MetricSPlat interface. (b) Propagation of result sets produced from similarity queries into FastMap projections. (c) Multivariate data visualization over subregions of the metric space. (d) Choosing new query centers, for new similarity queries.
 }
 \label{fig:systematization}
\end{figure*}

\noindent{The MetricSPlat system implements all the features needed in order to have a metric space in which similarity queries can be performed. Once the user defines a dataset, he/she can create a metric structure automatically, with one of several metrics structures and distance functions, see figure \ref{fig:systematization}(a). After this step, the analyst can select subregions of the data domain by choosing query centers of his/her interest and perform nearest neighbor or range similarity queries. The correspondent data elements are propagated to FastMap projections, see figure \ref{fig:systematization}(b), where it is possible to observe the region of the space that was selected. From the FastMap projections, it is possible to derive multi modal visualization techniques, as indicated in figure \ref{fig:systematization}(c). Finally, one can re-examine the visualizations and choose new query centers for further similarity queries, figure \ref{fig:systematization}(d). This process can repeat several times defining a multiple views environment where each view holds a different subregion of the data domain.}

In comparison to purely visual methods, in which the user can select subregions in a graphical data projection, the use of similarity queries: provides more precision on the task of selecting elements, is faster even compared to interactive methods, is immune to overlapping elements and carries the structural information embedded in the ensemble of a metric space.

\section{Experiments}
\label{sec:experiments}

\noindent{In this section we answer the questions raised in section \ref{sec:introduction}. We perform two experiments with two different datasets, each followed by conclusions about how the visual analytical process brought new possibilities for data investigation. Because the demonstration of our interactive system is not favored by the flat paper presentation, we have put the system fully operational at \url{http://gbdi.icmc.usp.br/~junio/MetricSPlat}, where it can be experimented with several datasets.}\\

\noindent{\bf Filtering out data with precision and accuracy}\\
\noindent{We have performed experiments over a dataset of agrometeorological data. The dataset has nine attributes: precipitation, maximum temperature, minimum temperature, normalized difference vegetation index (NDVI), water requirement satisfaction index (WRSI), average temperature, potential evapotranspiration (ETP), real evapotranspiration (ETR) and measured evapotranspiration (ETM) collected partly with remote sensors (satellite) and partly with {\it in locu} samples from sugar cane plantation regions in Brazil. The data was collected during 82 months from 2001 to 2007, so that each record corresponds to the data collected in a given month for one region. Since there are 5 regions (Araraquara, Araras, Jaboticabal, Ja\'u, and Ribeir\~ao Preto), there is total of 410 records. The goal of the data collection is to analyze the relationship between the weather (temperature and humidity) and the agricultural production.}

To test our methodology, we aim at creating visualizations that put together and that characterize regional weather conditions according to multiple simultaneous queries. Each query allows to choose a subregion of the data domain having a specific center to focus on. The goal is to visualize the subregions of interest out of the great data mass in dedicated visualization workspaces, where a simplified analytical process can proceed in a reduced graphical space. After an initial selection, the data is overviewed in a data projection and explored with multi modal multivariate visualization techniques, where patterns can be observed.

In figure \ref{fig:experiments1}, we illustrate three queries over the agrometeorological dataset. At the top of figure \ref{fig:experiments1}, it is possible to see the entire data domain in a FastMap projection. In this image, there are three regions of interest with their correspondent query centers. Figure \ref{fig:experiments1} also describes each of the three queries, Q1 and Q2. At the left side of the image, it is possible to see the initial Euclidean data selections in FastMap projections along with parallel coordinates. In Q1, the query center has id number 430, which corresponds to the region of {\it Ribeirao Preto} in the 18th month of observation along with 40 similar elements. Its Euclidean query returned a sparse subregion of similar elements. Its parallel coordinates, in turn, shows a distribution of elements tending to high evapotranspiration (the three right-most axes) and low vegetation rate (axis 3). The correspondent scatter plot, right, shows that this query center corresponds to a region of extreme temperatures in the dataset (right-top of scatter plot: average temperature x ETM) and that this region corresponds to a profile of regions that suffer from excessive sun radiation, defining a local correlation between high evapotranspiration and high temperature. The query has automatically selected for us the other regions that mostly correspond to that profile: {\it Jaboticabal} and {\it Jau}.

In query Q2, the center is element number 156 from the region of {\it Araras} in the 74th month of observation, surrounded by 40 similar elements. Its FastMap projection depicts three clusters, informing that there are three classes of records regarding the similarity to element 156. The correspondent parallel coordinates show that the profile of the most similar elements is that of high NDVI (axis 3), high WSRI (axis 4) and low evapotranspiration (the three right-most axes). Further, the table lens visualization shows an inverse proportionality between vegetation rate (column 3) and temperature (columns 2, 5 and 6). By examining the records presented in the FastMap visualization, we verified that this profile corresponds to the months when the sugar cane culture was in the middle of its annual cultivation cycle (high vegetation and mild temperatures), beginning of the cycle and, less intense, final of the cycle (low vegetation). The more close to the middle of the cycle, the higher is the amount of biomass in the crops, providing higher levels of vegetation. This observation explains the three clusters observed at the initial FastMap projection and, also, it explains why the third cluster is less observable. Had we defined a smaller number of elements, we would not have the third further most cluster with records corresponding to the end of the sugar cane cycle, when harvesting takes place.

\begin{center}
\includegraphics[width=\columnwidth]{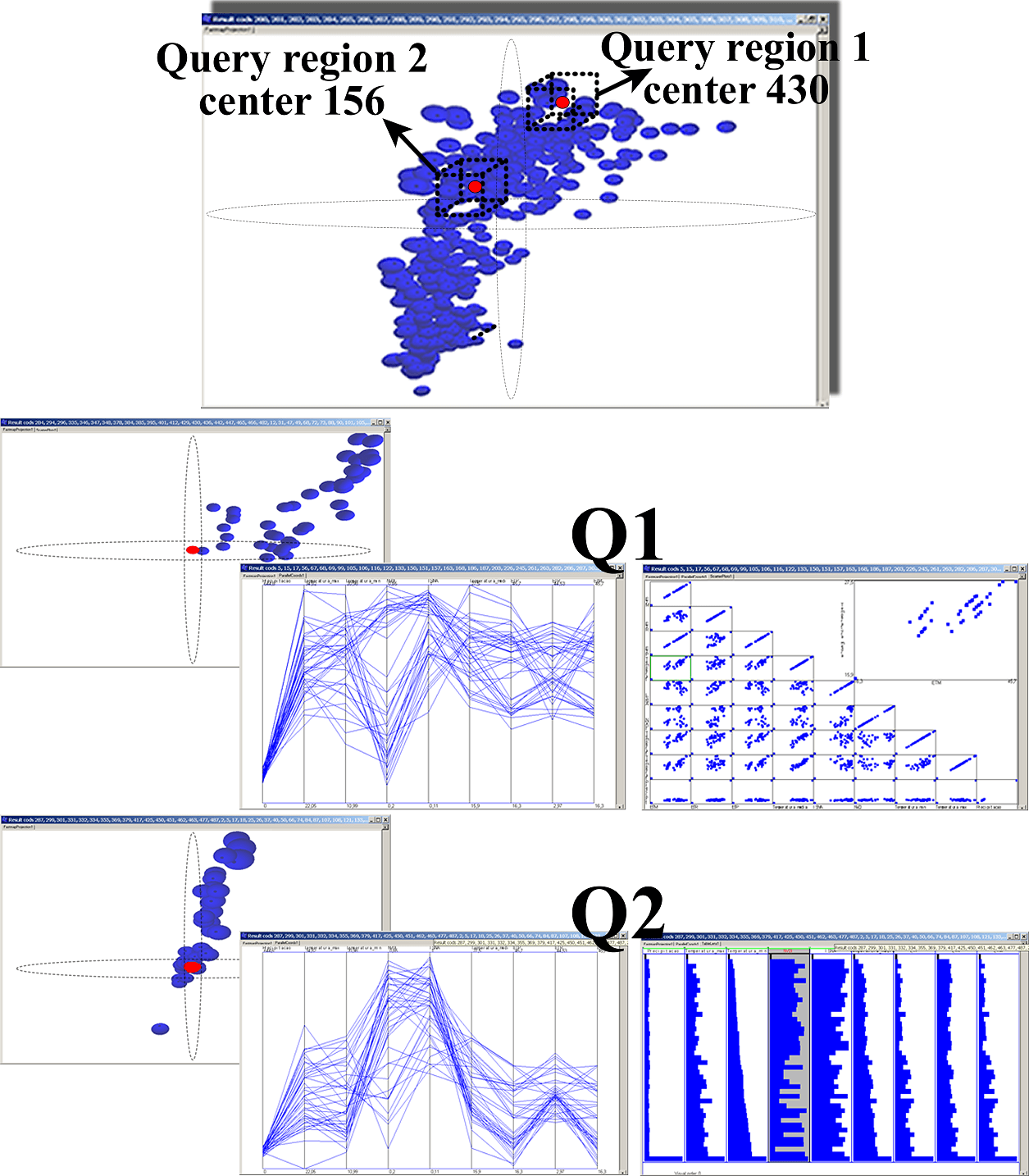}
\makeatletter\def\@captype{figure}\makeatother
\caption{Multiple views visualization of the agrometeorological dataset. At the top, an overview of the entire dataset. Following, queries Q1 and Q2, Euclidean distance, with two distinct query centers and correspondent multivariate visualizations, parallel coordinates, scatter plots and table lens.}
\label{fig:experiments1}
\end{center}

{\ }\\
\noindent{\bf Adding semantics in the visual data analysis process}\\
\noindent{Our second experiment uses the well-known semantic rich cars dataset (\url{http://stat.cmu.edu/datasets}). This dataset is broadly used due to its common sense domain, allowing readers to quickly understand what is being presented in only a few words. It has 8 dimensions and 406 data elements. The information in the dataset comes from car road tests performed by the Consumer Reports Magazine between 1971 and 1983. The attributes of the data are fuel performance in miles per U.S. gallon (MPG), number of cylinders in the engine (CYLINDERS), engine displacement in cubic inches (DISPLACEMENT), time to accelerate from 0 to 60 mph (ACCELERATION), output of the  engine in horsepower (HORSEPOWER), vehicle weight in U.S. pounds (WEIGHT), model year (YEAR), and origin of the car (ORIGIN). The last attribute organizes the dataset into three categories, American cars (1), Japanese cars (2) and European cars (3).}

In the following experiment, we want to show that with the aid of content-based data retrieval techniques, it is possible to add semantics to the visual analytical process. To do so we use two different distance functions: the Euclidean distance and the weighted Minkowski distance function with $p=4$ and two different sets of attribute weights. In the website of MetricSPlat, there is a video that shows how new distance functions can be added to the system in less than three minutes. Figure \ref{fig:experiments2} shows at its top an overview of the cars dataset in a FastMap projection using the Euclidean distance. 

Figure \ref{fig:experiments2}(Q1) refers to the data domain sub region around car number 4, an American car produced in 1970 that is considered to be a mostly non economic car, as all of its attributes deviate from the average toward values of high consume and high power. At Q1, left, the set of data items in the region around car number 4 corresponds to the 50 cars that more close present attributes similar to those of car number 4. The view achieved with the Euclidean distance, although showing a subset with distinguishable properties, does not present a highly characteristic silhouette in relation to the query center. Differently, the scene achieved with the weighted Minkowski distance, at the right, linearly weights the attributes so that the number of cylinders (CYLINDERS) and the weight of the cars (WEIGHT) have less impact on the placement of the data elements. The result is that even cars with significant differences in those attributes are put together in the spatial arrangement of the metric space and of the visualization. The correspondent visualization presents data elements whose attributes have smaller amplitudes than the Euclidean presentation, allowing the analyst to observe the cars that, even if semantically more powerful and heavier, have the other attributes similar to those of car number 4. That is, attributes CYLINDERS and WEIGHT, although sound, are not determinant for this specific subset.

Figure \ref{fig:experiments2}(Q2) presents the subregion of the data domain around car number 369, a Japanese car produced in 1981 with economic features as lower number of cylinders and high miles per gallon consume. At Q2, left, again we have the 50 cars closely similar to car 369 according to the Euclidean distance. At this second experiment, we noticed that linear weighting of attributes rarely produced result sets different from Q2 left visualization. Therefore, we have opted for an exponential weighting, so that the higher the difference between two data attributes, the much higher their distance in the metric space. We have weighted attributes ACCELERATION and HORSEPOWER, and at Q2, right, we can see a data subset that aggregates cars with similarity in those attributes. Semantically, what we see is the visualization of the cars with average economic profile (similar to car 369), at the same time that we can observe the local distribution of their acceleration and power. Hence, it is possible to have an idea of the cars that are mildly economic, with high and low power, and that are as close as possible to profile 369. Query Q2 shows how the use of intelligent distance functions can add semantics to the visual data analysis process, reducing the data domain, and the visual complexity, to reflect the analyst interests.

\noindent{\bf Making sense of metric spaces}\\
\noindent{In both experiments, it is possible to perceive the metric space defined by the data domain. This is possible due to the FastMap projection of the entire dataset along with the subregions selected via similarity queries. By joining the multiple visualizations interactively, the user can build a mental map of how the data elements are positioned in the space. This possibility, combined to metric space techniques, can lead to a potentially useful instrument for dealing with the challenges of content-based data retrieval.}

\begin{center}
\includegraphics[width=\columnwidth]{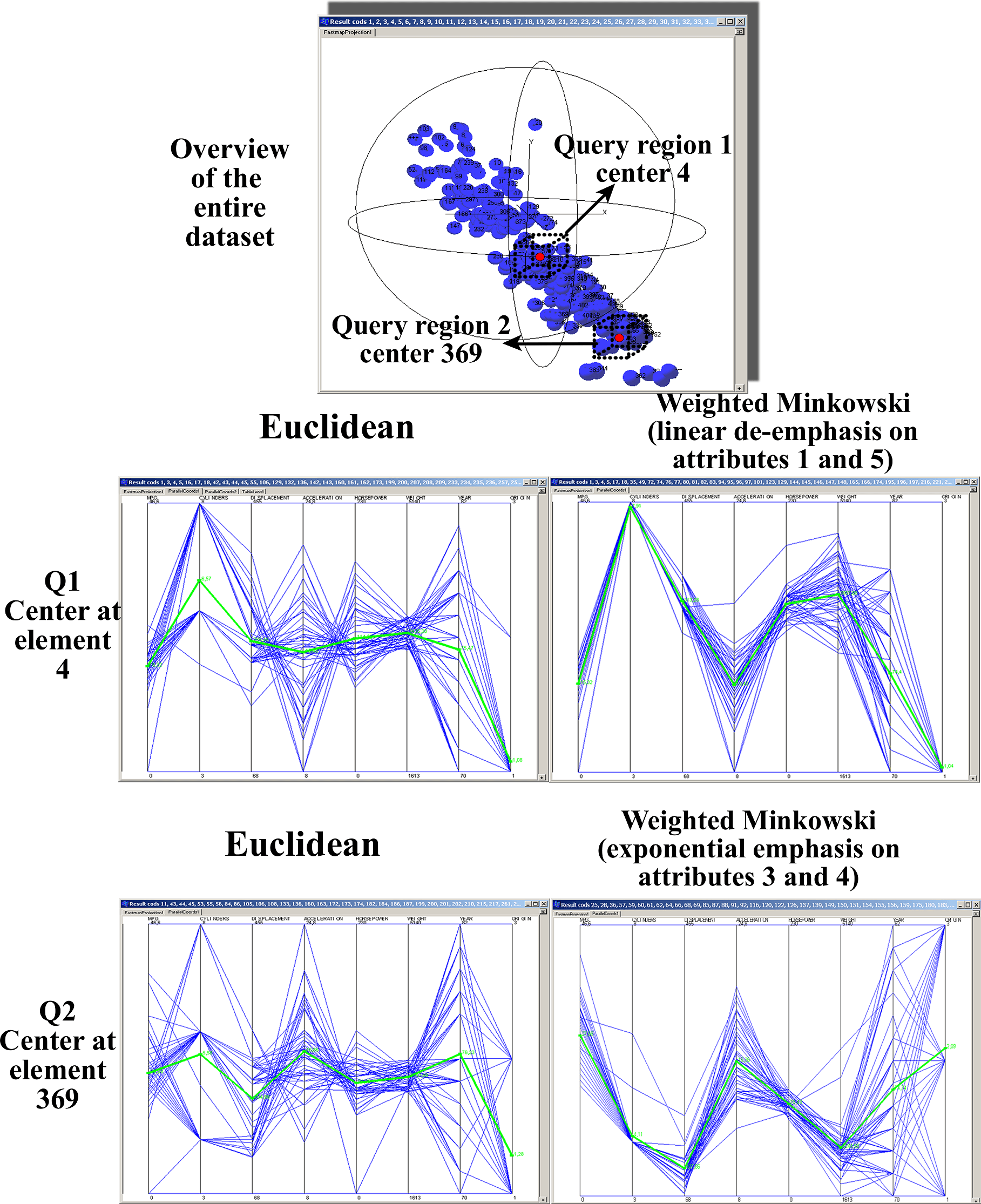}
\makeatletter\def\@captype{figure}\makeatother
\caption{Multiple views visualization of the cars dataset. At the top, an overview of the entire dataset. Following, queries Q1 and Q2 for Euclidean distance and for weighted Minkowski distance with two sets of attribute weights.}
\label{fig:experiments2}
\end{center}

\section{Conclusions and future work}
\label{sec:conclusions}

\noindent{In this work we benefit from two data management methodologies: content-based data retrieval (metric spaces) and visual data analysis. The combination of both methodologies has been used to answer three questions: how to efficiently filter out data for visual inspection, how to add semantics to the visual data analysis process, and how to quickly make sense of metric spaces. In order to answer these questions, we have built a system named MetricSPlat, which puts together a wide set of technologies related to similarity querying and multivariate data analysis. The system joins both expertises by using the FastMap multidimensional data projection. FastMap draws the metric space defined by the features, the distance functions and the metric data structure providing an intuitive spatial view of the data. This spatial view is the first step of our methodology, which is followed by multi modal multivariate data visualization. Complementing the system, we make use of multiple visualization workspaces presented simultaneously to the user. This feature is specially interest because it is useful to compare several spatial drawings of the data, enriching the analytical possibilities.}

We have demonstrated our system with two datasets. The first one is from agrometeorological research, and the second one, a classic dataset used in visualization demonstrations. We have showed the possibilities of efficiently filtering out the datasets with the aid of similarity queries. Also, we have presented tests on how to add semantics to the visual data analysis process by defining intelligent distance functions according to the analytical interests. Both experiments demonstrated the usefulness of the tool in depicting a metric space visually, instigating the user intuition about the data.

We conclude by stating that our work is the convergence of two methodologies related to the same big field of data management. Their convergence seems beneficial for both. Visual data analysis is improved by a filtering possibility more efficient than interactive filtering (brushing), as it can be more accurate and semantically rich. Content-based data retrieval is improved by the possibility of understanding the metric space that supports the similarity queries, allowing the fine tuning and advanced use of distance functions, features extraction, and metric data structures. We foresee future works following the benefits in both areas.\\

\section*{Aknowledgements}
This work is thank to Brazilian agencies Funda\c{c}\~ao de Amparo \`{a} Pesquisa do Estado de S\~ao Paulo (FAPESP), Conselho Nacional de Desenvolvimento Cient\'ifico e Tecnol\'ogico (CNPq), Coordena\c{c}\~ao de Aperfei\c{c}oamento de Pessoal de N\'ivel Superior (Capes), to Microsoft Research, and to STIC-AmSud cooperation program.

\bibliographystyle{IEEEtran}
\input{refs.bbl}

\end{document}